\documentclass[]{elsart3p}
\usepackage{graphicx}
\usepackage{amssymb}
\usepackage{bm}

\begin{document}

\begin{frontmatter}
\title{de Haas-van Alphen effect investigations of the electronic \\ structure of pure and aluminum-doped MgB$_2$}
\author[a]{A.~Carrington}
\author[a]{E.A.~Yelland}
\author[a]{J.D.~Fletcher}
\author[b]{J.R.~Cooper}
\address[a]{H.~H.~Wills Physics Laboratory,  University of Bristol, Bristol, BS8 1TL, England.}
\address[b]{Department of Physics,University of Cambridge, Madingley Road, Cambridge, CB3 0HE, England.}


\begin{abstract}
Understanding the superconducting properties of MgB$_2$ is based strongly on knowledge of its
electronic structure. In this paper we review experimental measurements of the Fermi surface
parameters of pure and Al-doped MgB$_2$ using the de Haas-van Alphen (dHvA) effect. In general, the
measurements are in excellent agreement with the theoretical predictions of the electronic
structure, including the strength of the electron-phonon coupling on each Fermi surface sheet. For
the Al doped samples, we are able to measure how the band structure changes with doping and again
these are in excellent agreement with calculations based on the virtual crystal approximation. We
also review work on the dHvA effect in the superconducting state.
\end{abstract}

\end{frontmatter}

\section{Introduction}
The physical properties of MgB$_2$ are strongly linked to its unusual electronic structure. This
has been extensively investigated theoretically and has led to a detailed microscopic understanding
of the physics of this material. Experimentally, there are many probes which are sensitive to
various aspects of the anisotropic electronic structure, although there are relatively few which
can determine the most fundamental parameters accurately.  For example, measurements of the
resistivity can be used to estimate the average mean-free-paths $\ell$ along the two principal
crystallographic directions ($a$ and $c$), but cannot provide $\bm{k}$-resolved information.

Knowledge of these $\bm{k}$-dependent parameters and how they vary with doping is important for
understanding the superconducting properties of MgB$_2$, such as $T_c$ and the upper critical field
$H_{c2}$. In particular, a knowledge of how a certain dopant changes the band filling and the
scattering rates on the various Fermi surface sheets can help in optimizing the superconducting
properties for applications. For example, in order to raise $H_{c2}$ it is necessary to lower the
mean-free-path, especially on the $\sigma$ sheets of the Fermi surface, which have a large
superconducting gap, without depressing $T_c$ too much by mixing in states from other parts of the
Fermi surface where the gap is smaller.

The de Haas-van Alphen (dHvA) effect, is a powerful method for determining the Fermi surface
parameters of metals, providing information about the size and shape of the Fermi surface sheets,
and the scattering rates and quasiparticle effective masses on these sheets \cite{Shoenberg}.  The
data are usually interpreted using Lifshitz-Kosevich (LK) theory \cite{Shoenberg}. The main
experimental constraint comes from the fact that dHvA oscillations are only observable if the
quasiparticles complete a significant fraction of a cyclotron orbit before being scattered. In
practice, high magnetic fields and reasonably pure samples with long mean-free-paths are needed.
This limits the study of doping effects on the electronic structure, since doping is usually
achieved by partial atomic substitution which inevitably increases scattering and therefore
decreases $\ell$. The amplitude of the dHvA signal is reduced from its value in a perfectly clean
sample by the Dingle factor \cite{Shoenberg}, $R_D =
\exp(-\frac{\pi}{\omega_c\tau})=\exp(-\frac{\pi \hbar k_F}{eB\ell})$. The damping is increased with
shorter $\ell$ and increased average Fermi wavevector ($k_F$) of the orbit (or equivalently as the
product of the cyclotron frequency $\omega_c$ and the scattering time $\tau$ decreases). This can
be offset, to some extent, by making measurements at very high magnetic fields $B$, although in
practice, we are still limited to dopant levels of a few atomic percent where the increase in
scattering is not too severe.

In this paper we will review studies of the dHvA effect in both pure and aluminum doped MgB$_2$,
and information obtained about the electronic structure and superconducting properties.

\section{Electronic structure}

\begin{figure}
\begin{center}
\includegraphics[width=75mm]{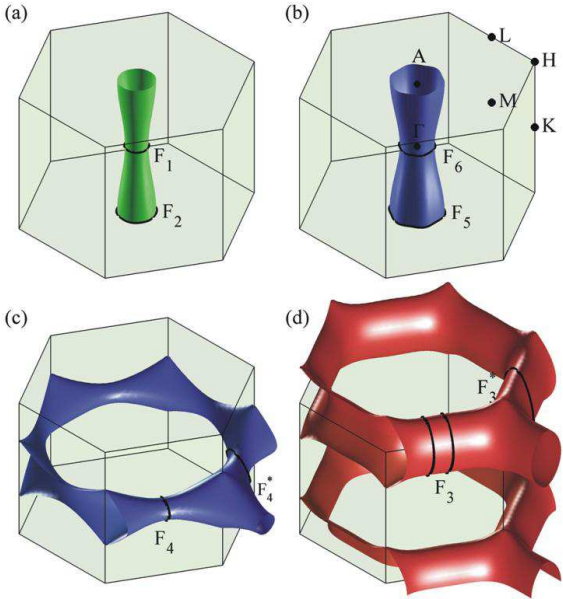}
\end{center}
\caption{(color) Calculated Fermi surface of MgB$_2$, with possible dHvA extremal orbits (for
frequencies $<$10~kT) indicated. Panels (a)-(d) show the $\sigma$-light hole, $\sigma$-heavy hole,
$\pi$-hole and $\pi$-electron bands respectively.} \label{fspic}
\end{figure}

The interpretation of dHvA data is considerably aided by detailed comparison with band structure
calculations. The first calculations of the band structure of MgB$_2$ were reported long before the
discovery of superconductivity in MgB$_2$ (see Ref.\ \cite{MazinA03}).  Shortly after the discovery
of superconductivity, Kortus \textit{et al.} \cite{KortusMBAB01} reported detailed calculations of
the Fermi surface which specifically address the issue of the origin of the superconductivity.
Since then there have been a large number of other reports which have confirmed and added to this
picture (see for example, Refs. \cite{AnP01,ChoiRSCL02b,LiuMK01,Harima02,RosnerAPD02}).

Generally, the electronic orbits which give rise to dHvA oscillations are located at the points
where there is a local minimum or maximum in the Fermi surface cross-sectional area along the
direction of the applied magnetic field. The observed dHvA frequency is proportional to this
`extremal area'. In most cases these occur at obvious symmetry points in the reciprocal unit cell,
although the origin of some orbits is more subtle. Several papers
\cite{Harima02,MazinK02,RosnerAPD02} have reported calculations of the dHvA extremal cross-sections
and effective masses, and generally these are in good agreement with each other.

In Fig.\ \ref{fspic}, we show the Fermi surface of MgB$_2$, calculated using the Wien2K package
\cite{wien2k} (see later). We have indicated the locations of the predicted dHvA orbits
\cite{Harima02,MazinK02,RosnerAPD02}.  These have been labelled $F_1 \ldots F_6$ in accordance with
previous publications \cite{YellandCCHMLYT02,CooperCMYHBTLKK03,CarringtonMCBHYLYTKK03,Isshiki2005}.

\section{Experimental details}
The Landau level quantization of the energy levels of a metal in a strong magnetic field leads to
oscillations of the density of states as a function of the magnetic field and thus to corresponding
oscillations in the physical properties  \cite{Shoenberg}. The most popular technique is to measure
the oscillations in the differential magnetic susceptibility of the sample. In practice, the
sensitivity of this technique is limited by the volume of the sample. In the case of MgB$_2$, the
available highest quality single crystals are relatively small ($<$0.05mm$^3$), so better signal to
noise can be obtained by measuring the oscillations in the torque, using a piezo-resistive
cantilever of the type developed for atomic force microscopy \cite{RosselBZHWK96}. All the results
we will show here were obtained by this method.  The details of the method are described in Ref.\
\cite{CooperCMYHBTLKK03}.

\section{Results for pure MgB$_2$}
The first report of dHvA oscillations in MgB$_2$ was by Yelland \textit{et al.}
\cite{YellandCCHMLYT02}.  In this report, which used magnetic fields up to 18~T, three dHvA
frequencies ($F_1$, $F_2$, and $F_3$) were observed. $F_1$ and $F_2$ correspond to the extrema of
the smaller (light hole) $\sigma$ tube, and $F_3$ is located on the electron-like $\pi$ sheet. In a
later study on crystals produced from isotopically pure $^{10}$B, Carrington \textit{et
al.}\cite{CarringtonMCBHYLYTKK03} observed oscillations from all four sheets of the Fermi surface
($F_1 \ldots F_6$ in Fig.\ \ref{fspic}), using fields up to 33 T at the Tallahassee high magnetic
field laboratory.

\begin{figure}
\begin{center}
\includegraphics[width=7.0cm,keepaspectratio=true]{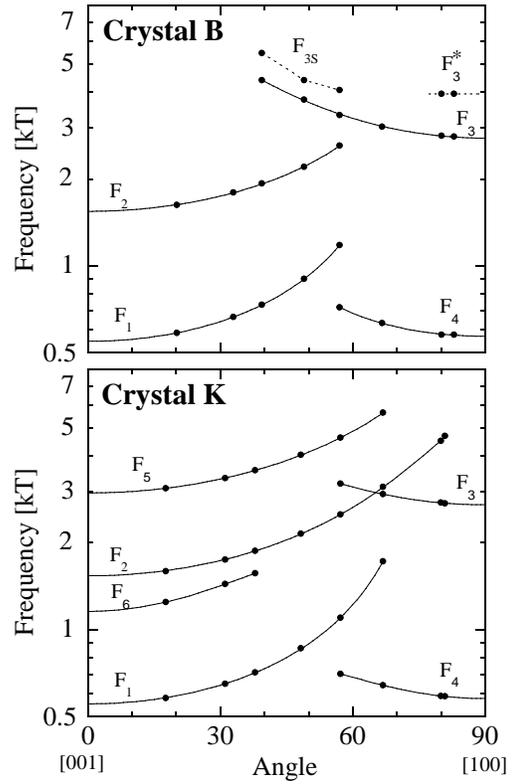}
\end{center}
 \caption{Observed frequencies versus field angle as the samples were rotated from $H \|c$ to (approximately) $H \|a$.
 The solid lines are polynomial fits , and the dotted lines are guides to the eye.} \label{rotplot}
\end{figure}

The observed dHvA frequencies as a function of field angle are shown in Fig.\ \ref{rotplot}.
Results for two different samples are shown.  Sample B was grown from natural (mixed isotope) boron
whereas sample K was the pure $^{10}$B sample. In crystal K signals from all four sheets of the
Fermi surface were observed, however, in crystal B no oscillations arising from the larger (heavy
hole) $\sigma$ sheet were observed even in these higher fields. This is almost certainly because of
the longer mean-free-paths on the $\sigma$ sheets of crystal K relative to those in B (see Table
\ref{tablesum}). Note the opposite trend of $\ell$ on the $\pi$ sheets of the two crystals. In
general, we find that the mean-free-paths on the $\sigma$ sheet are not well correlated with those
on the $\pi$ sheets. This will be discussed later.

The dHvA frequencies, extrapolated by polynomial fits to the symmetry points $F^0$, are given in
Table \ref{tablesum}. We find very good reproducibility between samples from different sources, and
the dHvA frequencies agree to within 30~T or 0.06\% of the basal area of the first Brillouin zone
($\frac{\hbar}{2\pi e}\frac{8\pi^2}{\sqrt{3}a^2}=50.2$~kT).  This is evidence against there being
any significant Mg deficiency in the samples. In general, the agreement between the measured $F^0$
values and those predicted by theory is very good although not perfect. To investigate the size of
the discrepancy we apply a rigid shift to the calculated energies of the $\sigma$ and $\pi$ bands.
Since the dHvA band mass is defined as $m_B=\frac{\hbar^2}{2\pi}\frac{\partial A}{\partial E}$, the
necessary band shifts are given by $\Delta E = \frac{\hbar e }{m_B} \Delta F$ (where $\Delta F =
F_{\rm LDA}-F_{\rm exp}$). For the $\sigma$ sheet orbits (1,2,5,6) the average shift is $83\pm 4$
meV, whereas for the $\pi$ sheet orbits (3,4) it is $-61\pm 5$ meV.
 The volumes of the $c$-axis tubes are proportional to the average of the two extremal areas
\cite{RosnerAPD02} and these are both $\sim 16$\% smaller than the calculations
\cite{MazinAJDKGKV02}, implying a corresponding difference in the number of holes in these two
tubes \cite{RosnerAPD02}.

\begin{figure}
\begin{center}
\includegraphics[width=7.0cm,keepaspectratio=true]{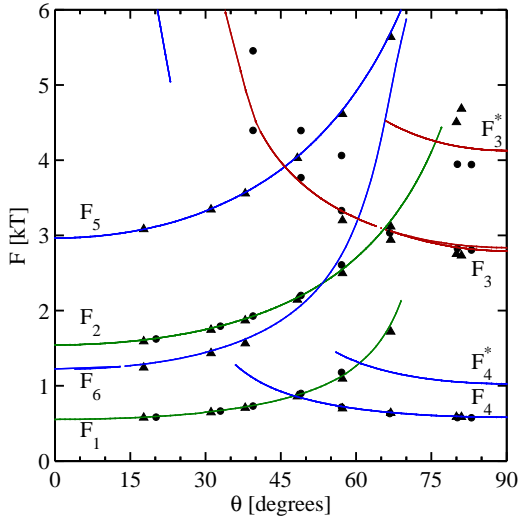}
\end{center}
\caption{(color) Comparison of the observed dHvA frequencies with extremal FS areas extracted from
the calculated band structure ($\blacktriangle$=crystal K, $\bullet$=crystal B). Calculated
frequencies are for an in-plane angle $\phi=12^\circ$, to match crystal B.} \label{rotplotcompWien}
\end{figure}

In Fig.\ \ref{rotplotcompWien} we compare the frequencies observed experimentally with the
predictions based on the LDA band structure. Here we have re-calculated the band structure using
the Wien2K package \cite{wien2k} (using parameters similar to those of Kortus \textit{et al.}
\cite{KortusMBAB01}) on a very dense $k$-mesh (95000 points in the full Brillouin zone) and
extracted the dHvA parameters using an automated numerical search algorithm. The bands have been
rigidly shifted to give the best agreement at the symmetry points as described above. The agreement
is excellent, showing that the LDA calculations (with an applied small rigid shift) give the
correct Fermi surface.

\begin{table*}
\caption{Summary of dHvA parameters for both samples (K and B) along with the theoretical
predictions (Th) \cite{MazinK02}}
\begin{center}
\begin{tabular}{cc|ccccccccc|cccc}
\hline\hline
&&\multicolumn{9}{c}{\bf Crystal K}&\multicolumn{4}{|c}{\bf Crystal B}\\
 Orbit&&$~F^0_{\rm exp~}$&$~F^{~}_{\rm Th}~$&$\Delta F$&$~\Delta E$&~$\ell$~&~$m^*_{\rm exp}$~&~$m_{_{\rm
band}}$~&$\lambda^{\rm ep}_{\rm exp}$&~$\lambda^{\rm ep}_{\rm Th}$~&~$F^0_{\rm exp~}$&~$\ell$~&~$m^*_{\rm exp}$~&$\lambda^{\rm ep}_{\rm exp}$\\
& &[T]  & [T]& [T] & [meV] & [\AA] & [$m_e$] &[$m_e$]&~&~&[T]&[\AA]&[$m_e$]&~\\
\hline
1&$\sigma$&551&730&$+179$&$+83$&550&$0.548\pm0.02$&0.251&$1.18\pm0.1$&1.25&546&380&$0.553\pm0.01$&$1.20\pm0.04$\\

2&$\sigma$&1534&1756&$+222$&$+82$&900&$0.610\pm0.01$&0.312&$0.96\pm0.03$&1.25&1533&580&$0.648\pm0.01$&$1.08\pm0.03$\\

3&$\pi$&2705&2889&$+184$&$-67$&570&0.439$\pm0.01$&0.315&$0.40\pm0.03$&0.47&2685&680&$0.441\pm0.01$&$0.40\pm0.03$\\

4&$\pi$&576&458&$-118$&$-56$&$-$&$0.31\pm0.05$&0.246&$0.31\pm0.1$&0.43 &553&$-$&$0.35\pm0.02$&$0.42\pm0.08$\\

5&$\sigma$&2971&3393&$+422$&$+79$&390&$1.18\pm0.04$&0.618&$0.91\pm0.07$&1.16&$-$&$-$&$-$&$-$ \\

6&$\sigma$&1180&1589&$+409$&$+87$&$-$&$1.2\pm0.1$&0.543&$1.2\pm0.2$&1.16&$-$&$-$&$-$&$-$ \\

\hline\hline
\end{tabular}
\end{center}
\label{tablesum}
\end{table*}

In MgB$_2$ the dominant source of mass enhancement is the electron-phonon interaction
\cite{ChoiRSCL02b,LiuMK01,MazinK02}. If we assume this is the only source of enhancement we can
estimate the electron-phonon coupling constants $\lambda^{\rm ep}$, from the measured quasiparticle
effective masses $m^*$, $\lambda^{\rm ep}=m^*/m_B -1$. The results (Table \ref{tablesum}) show that
the values of $\lambda^{\rm ep}$ on both the $\sigma$ sheets are approximately a factor three
larger than those on the $\pi$ sheets, in agreement with theoretical predictions. This is a key
factor in the multi-band nature of superconductivity in MgB$_2$, and results in a much larger
superconducting gap on the $\sigma$ sheets than on the $\pi$ sheets.

The experimental values of $\lambda^{\rm ep}$ are generally slightly smaller than the theoretical
ones (even when the $\sigma$ and $\pi$ band shifts, mentioned above, are taken into account
\cite{CarringtonMCBHYLYTKK03}. One possible reason for this is that these theoretical values do not
take into account phonon anharmonicity, which has been shown to reduce the average value of
$\lambda$ by around 20\%\ \cite{ChoiRSCL02}.

In sample B, two additional dHvA frequencies were observed which we did not see in crystal K (these
are labelled $F_3^*$ and $F_{3s}$).  When the field is aligned in-plane, close to the $a$-axis
[$100$], the orbit $F_3$ can be seen, but in addition there are also two orbits in the
symmetrically equivalent section of Fermi surface running along the [$110$] and [$0\bar{1}0$]
directions, inclined at 60$^\circ$ to the main $F_3$ orbit. The origin of $F_{3s}$ has not yet been
conclusively identified.

As far as we know, there has been only one other report of dHvA measurements in MgB$_2$ by another
group. Isshiki \textit{et al.} also used a piezoresistive cantilever method to measure crystals
grown by an encapsulation technique \cite{Isshiki2005}.  Their results are very similar to those
shown above for crystal B (Fig.\ \ref{rotplot}). They observe all the same frequencies (including
$F_3^*$ and $F_{3s}$ in Fig.\ \ref{rotplot}), but do not see any signals from the heavy hole
$\sigma$ sheet ($F_5$ and $F_6$), most likely because of a lower mean-free-path in these samples.
The frequencies at the symmetry points ($F^0_N$) are within 20T of those in Table \ref{tablesum}
except $F_4$ which is 46~T lower \cite{aoki2006}.

\section{Results for Al-doped MgB$_2$}

Studies of the effect of atomic substitutions in MgB$_2$ are important for applications and may
provide a route to improve its superconducting properties. It also allows us to test theoretical
understanding of the material. The two elements which substitute most readily in MgB$_2$ are Al and
C which replace Mg and B respectively, to give Mg$_{1-x}$Al$_x$(B$_{1-y}$C$_y$)$_2$. Both dopants
add electrons to the material, and cause $T_c$ to decrease in a similar way \cite{KortusDKG05}.
However, the effect of each dopant on $H_{c2}$ and resistivity is very different. Al doping gives a
moderate increase in the residual resistivity and causes only a small change in $H_{c2}$ and its
anisotropy. On the other hand, carbon doping dramatically increases both the resistivity and
$H_{c2}$\ \cite{KazakovPRMZJSBK05,KarpinskiZSKBRPJMWGDUS05}. This is understandable as the $\sigma$
band has most weight on the boron plane, whereas the $\pi$ band has weight on both the boron and
magnesium planes. It is therefore to be expected that replacing boron with carbon should strongly
increase the scattering rates on both the $\sigma$ and $\pi$ bands, but replacing Mg with Al would
mainly affect the $\pi$ band.

dHvA studies can test theoretical predictions by measuring sheet-specific scattering rates and by
determining quantitatively how doping changes the volumes (or cross-sections) of Fermi surface
sheets and the effective masses of quasiparticles. The latter can indicate the doping-dependence of
the electron-phonon coupling constants.

Attempts were made to detect dHvA oscillations in both Al and C doped samples with similarly
reduced $T_c$ \cite{CarringtonFCTBZKKCK05}. The C doped samples had  $T_c$ values of 35.7~K and
34.5~K, and nominal C contents of $y=3\%$ and $y=4\%$ respectively. Their residual resistivities
$\rho(T_c)$ were 11$\mu\Omega$\,cm and 30~$\mu\Omega$\,cm, substantially larger than for pure
MgB$_2$ which typically has $\rho(T_c)\sim 1\mu\Omega$\,cm. No dHvA signals from either the
$\sigma$ or $\pi$ sheets were seen in either C doped sample.  The two Al doped samples (AN215 and
AN217) both had $T_c\simeq 33.6$~K (as determined by heat capacity measurements), and Al contents
of $x=7.9\pm0.4$\% and $x=7.4\pm0.4$\% (as determined from their \textit{c}-axis lattice parameters
\cite{KarpinskiKJZAPW04}). The residual resistivity of the Al doped samples was found to be $\sim
4.5~\mu\Omega$\,cm.

\begin{figure}
\begin{center}
\includegraphics[width=6.0cm,keepaspectratio=true]{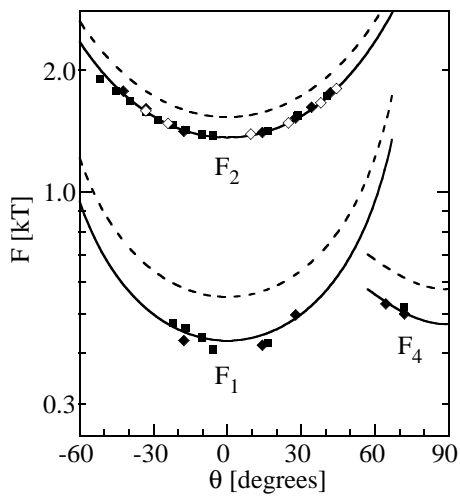}
\end{center}
\caption{Observed dHvA frequencies versus angle for the two Al doped samples AN215
($\blacksquare$,$\blacklozenge$) and AN217 ($\lozenge$). Dashed lines are $F(\theta)$ for pure
MgB$_2$, solid lines are pure $F(\theta)$ data scaled to fit the Al doped data.} \label{figrotAl}
\end{figure}

In Fig.\ \ref{figrotAl} we show the observed dHvA frequencies versus field angle, $\theta$.  Only
signals arising from the small (light hole) $\sigma$ sheet were observed. This sheet has the
smallest cross-section (or $k_F$) so the increased scattering (or decreased $\ell$) has the
smallest effect on the Dingle damping term. The values of $F_1^0$ and $F_2^0$ are the same in both
Al doped crystals, and are $10-20\%$ smaller than the corresponding values for pure MgB$_2$ (see
Table \ref{freqtable}). Studies on many different samples of pure MgB$_2$ (from different sources)
have revealed remarkably reproducible dHvA frequencies and so the measured reduction in
cross-section is significant.  Within a rigid band shift model our results imply that at this
doping the $\sigma$ bands have shifted to lower energy by $\sim$70~meV relative to the experimental
results for the pure samples.  From the calculated volume of these sheets we find that there are
$16\pm2$\% fewer holes in 7\% Al doped samples compared to pure MgB$_2$.

This simple analysis was improved considerably by J.\ Kortus
\cite{KortusDKG05,CarringtonFCTBZKKCK05} who calculated the effect of Al doping on the band
structure using the virtual crystal approximation (VCA). Here the effect of doping with Al is
simulated by replacing the Mg atom with a virtual atom with charge $Z=xZ_{\rm AL}-(1-x)Z_{\rm Mg}$.
These calculations \cite{CarringtonFCTBZKKCK05} give the following dependence of dHvA frequencies
on Al content $x$, $F_1=540-1820x$, $F_2=1530-2050x$ (here the bands have been rigidly shifted to
give agreement with the experimental data at $x=0$, i.e. for pure MgB$_2$). The observed
frequencies ($F_1$ and $F_2$) correspond to a doping of $x=7.5\pm 1$\% and $8.4\pm1$\%
respectively. These values compare favorably to the $x$ values deduced from the $c$-axis lattice
constant of the same crystal ($\overline{x}=7.7\pm0.4$\%), and so we conclude that the VCA results
accurately describe the band filling effect of Al doping in MgB$_2$.

\begin{table}
\caption{Calculated (LDA)\cite{MazinK02} and measured dHvA frequencies [$F^0\equiv F(\theta=0)$]
for pure \cite{CarringtonMCBHYLYTKK03} and Al doped MgB$_2$. $\Delta E$ is the rigid band shift
needed to bring the theoretical values in line with experiment.} \label{freqtable}
\begin{tabular}{ccccccccc}
\hline\hline
    &&  LDA&  \multicolumn{2}{c}{Pure} & \multicolumn{2}{c}{AN215}  & AN217 \\
Orbit    &  &$F^0$ &  $F^0$ &$ \Delta E$ &  $F^0$ &  $\Delta E$ &$F^0$ \\
    &&[T]&[T]&[meV]&[T]&[meV]&[T]\\
 \hline
$F_1$ & $\sigma$&  730 &  546$\pm$20 & 85 &  410$\pm$20 &  148     &$\cdots$ \\
$F_2$ & $\sigma$& 1756 & 1533$\pm$20 & 93 & 1360$\pm$20 &  147     & 1360$\pm$20 \\
$F_3$ & $\pi$& 2889 & 2685$\pm$20 &-75 & $\cdots$    & $\cdots$ &$\cdots$\\
$F_4$ &  $\pi$&  458 &  553$\pm$10 &-45 &  480$\pm$40 &-10       & $\cdots$ \\
\hline\hline
\end{tabular}
\end{table}

The VCA calculations are also able to predict the decrease in $T_c$ expected from the doping. For
each doping the density of states, phonon frequencies and electron-phonon coupling parameters were
calculated.   Close to $x=0$ the calculations \cite{KortusDKG05} predict $dT_c/dx=-0.50$ K/\%, and
hence for $x=7.7$\% we expect a $T_c$ reduction of $4.0\pm0.3$~K. The actual $T_c$ reduction in our
samples was 3.9~K, in excellent agreement with the calculations.  Note that these calculations do
not include the effect of the increased scattering, so it would appear that, at least for this
doping level, the scattering has a minimal effect on $T_c$.

A further test is to look at the decrease in the measured electron-phonon coupling strength with
doping.  For the Al doped samples, the only orbit where we have sufficient accuracy to make a
meaningful comparison is $F_2$.  For this orbit we find that $\lambda_{\rm ep}$  is $10\pm5$\%
smaller than for the pure case (the quoted error includes the measured variation in $\lambda_{\rm
ep}$ between different pure samples) \cite{CarringtonFCTBZKKCK05}. In calculating this we have
taken into account the small change in the band mass with doping. Theoretically, it is found that
$\lambda_{\rm ep}$ does not vary substantially within each $\sigma$ sheet and only varies by $\sim
7\%$ between the two $\sigma$ sheets \cite{MazinK02}, so the observed reduction in $\lambda_{\rm
ep}$ measured on orbit $F_2$ is expected to be representative of the changes on the other $\sigma$
sheet.  For $x=7.7$\%, the VCA calculation \cite{KortusDKG05,CarringtonFCTBZKKCK05} gives
$\lambda_{\rm ep}^\sigma=1.16$, which is $\sim 5\%$ smaller than for $x=0$. This is consistent with
our observations.

The mean-free-path, estimated from the Dingle term $R_D$, was $\ell=270$~\AA\ for orbit $F_2$. This
is approximately half of that found for pure samples of MgB$_2$ (grown with natural mixed-isotope
boron).  The signal from the electron-like $\pi$ sheet (frequency $F_3$) is normally large for
fields oriented near the plane, \cite{YellandCCHMLYT02,CarringtonMCBHYLYTKK03} and as we have not
observed this orbit in either Al doped sample we conclude that the mean-free-path on this sheet
must be reduced by a somewhat larger factor ($\gtrsim 3$).

Using this information we can make some simple estimates of the effect of Al doping on $H_{c2}$. In
the clean limit, $\mu_0H_{c2}=\Phi_0/(2\pi\xi^2)$, where $\xi$ can be estimated from the BCS
coherence length $\xi_0= \hbar v_F/(\pi \Delta)$. At low temperature the $\sigma$ sheets dominate
\cite{MiranovicMK03,GolubovK03,PuttiFMPTMPDGS05,AngstBWC05} and the anisotropy of $H_{c2}$
($\gamma_{H_{c2}}$) is determined by the anisotropy of $v_F$ on the $\sigma$ sheets, i.e.,
$\gamma_{H_{c2}}=H_{c2}^{\|a}/H_{c2}^{\|c}=\sqrt{\langle v^2_{F,a}\rangle/\langle
v^2_{F,c}\rangle}$. In the dirty limit, the same result is found \cite{GolubovK03} provided the
scattering is isotropic.  LDA calculations show that for $x\lesssim 0.3$,
\cite{PuttiFMPTMPDGS05,ProfetaCM03}
\begin{equation}
\sqrt{\langle v^2_{F,a}\rangle/\langle v^2_{F,c}\rangle}\simeq 5.9-7.5x. \label{gammavsx}
\end{equation}
Experimentally, we find for pure MgB$_2$, $\mu_0H_{c2}^{\|c}=3.5\pm0.2$T and $\gamma_{H_{c2}}=5.5
\pm 0.2$, at $T\simeq0.3$K, whereas for our Al doped samples ($x=7$\%),
$\mu_0H_{c2}^{\|c}=3.5\pm0.1$T and $\gamma_{H_{c2}}=4.6 \pm 0.2$. These values are in line with
other studies
\cite{RydhWKKCBLKMKKKLL04,PuttiFMPTMPDGS05,AngstBWC05,FletcherCTKK05,KarpinskiZSKBRPJMWGDUS05,KleinLMMSHSKKLLL06}.
Eq.\ (\ref{gammavsx}) slightly overestimates the anisotropy; however, the agreement improves if we
take into account the fact that pure MgB$_2$ has smaller $\sigma$ tubes compared to the LDA
calculations. The hole doping of the $\sigma$ sheets for pure MgB$_2$ is roughly equivalent to that
in the LDA calculations with $x\simeq 11$\%, so the predicted $\gamma_{H_{c2}}$ has a value of 5.2
for the pure case and 4.6 for the Al doped ($x$=7\%) case. These latter values are in good
agreement with experiment.  In addition, using the values of $\Delta$ and $v_F$ for the $\sigma$
sheets, $\xi_0\simeq 130$\AA. This is less than $\ell$ on these sheets for both the pure and
$x$=7\% samples and hence the lack of significant change in $\mu_0 H_{c2}(T=0)$ is also
understandable. In other words the $\sigma$ bands for both the 0\% and 7\% Al-doped samples are
moderately clean ($\xi_0\lesssim \ell$).

\section{Angle dependence of dHvA amplitude}

The angular dependence of the dHvA amplitude is often used to extract information about the
strength of electron-electron interactions. The Lifshitz-Kosevich expression describing dHvA
oscillations in a paramagnetic metal contains a factor $R_S=\cos[\frac{1}{2}\pi g
(1+S)m_\mathrm{B}/m_e]$ which arises from the phase difference between dHvA oscillations from the
spin-split Fermi surfaces \cite{Shoenberg}. Here, $S$ is the Stoner enhancement due to the
electron-electron interactions that enhance the Pauli spin susceptibility. At certain field angles,
oscillations from the spin-split Fermi surface sheets may destructively interfere; $R_S$ then
vanishes and a `spin-zero' is observed in the dHvA amplitude. Measurement of the field-angle
$\theta_\mathrm{sz}$ for a spin-zero can then be used to determine the Stoner enhancement
$S=(2/g)(n+\frac{1}{2})m_e/m_\mathrm{B}(\theta_\mathrm{sz})-1$, as long as the band-mass
$m_\mathrm{B}(\theta)$, $n$ and $g$ are known. $n$ can often be deduced from the frequency of the
`spin-zeros' as a function of angle; $m_\mathrm{B}(\theta)$ is taken from LDA band structure
calculations.

\begin{figure}
\begin{center}
\includegraphics[width=7.5cm,keepaspectratio=true]{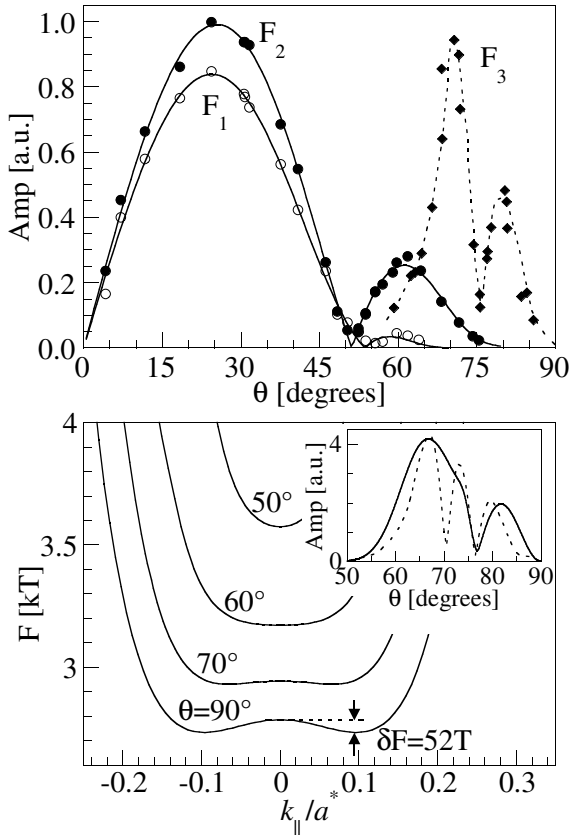}
\end{center}
\caption{Top panel: dHvA torque amplitude versus angle for frequencies $F_1$, $F_2$ and $F_3$
($17.8<B<18.0~$\,T). For frequencies $F_{1}$ and $F_{2}$ the solid lines are fits to the LK
expression including the spin-splitting factor $R_S$. For $F_3$ the dotted line is a guide to the
eye. Bottom panel: $F(k_\parallel)=\hbar A(k_\parallel)/2\pi e$ for certain field angles, from LDA
calculations. For fields close to the boron plane, both minimal and maximal cross-sections are
present, but for $\theta<65^\circ$ there is only one extremal orbit. Inset: the amplitude of $F_3$
fitted using a two frequency model (solid line) and calculated from the LDA band-structure (dashed
line, see text).} \label{figamp}
\end{figure}

Fig.\ \ref{figamp} shows the dHvA amplitude versus field angle for orbits $F_1$, $F_2$ and $F_3$.
The dips in the amplitudes of $F_1$ and $F_2$ are due to spin-zeros \cite{CarringtonMCBHYLYTKK03};
fits to the LK expression including $R_\mathrm{S}$ (solid line) are in excellent agreement with the
data. The Stoner enhancements are estimated to be $S(\mathrm{F_1})=0.10$ and $S(\mathrm{F_2})=0.14$
\cite{stonernote}. These are approximately a factor 2 lower than the values from band structure
calculations \cite{MazinK02}. The dHvA values are more consistent with conduction electron spin
resonance measurements of the spin susceptibility \cite{SimonJFMGFPBLKC01}, which imply no
enhancement ($S=0$) to within the experimental error of $\pm 15$\%.

A sharp minimum in the amplitude of the $F_3$ orbit on the $\pi$ band sheet was also observed (at
$\theta\approx 74^\circ$, $\phi=0^\circ$). This was initially interpreted as a spin-zero
\cite{YellandCCHMLYT02,CooperCMYHBTLKK03,CarringtonMCBHYLYTKK03}, although this required an
anomalously large Stoner enhancement. A later detailed dHvA study \cite{CarringtonFH05} revealed
that the dip is in fact due to a small warping of the nearly cylindrical part of this sheet which
can be seen in LDA calculations (see Fig.\ \ref{figamp}). This warping means that there are two
extremal dHvA orbits which have very similar frequencies for $\theta>74^\circ$. For lower $\theta$
only a single frequency is observed. Fits of the observed angular dependence using a model with two
discrete frequencies separated by $\Delta F \simeq 1.4 (\theta-70^\circ)$~T  can reproduce a sharp
minimum (see Fig.\ \ref{figamp}), but are not completely satisfactory \cite{CarringtonFH05}. It is
well known that the extremal orbit approximation used in the LK formula breaks down when there are
significant regions of the Fermi surface over which $F(k_\parallel)$ varies by $\lesssim B$, so
that changes in the dHvA phase cannot be treated as $\gg2\pi$. The effect is predominant in metals
with a quasi-two-dimensional Fermi surface \cite{YoshidaMSMIOBHMAS99,BergemannJMNM00}, but it is
also significant in the nearly cylindrical sections of the electron-like $\pi$ band Fermi surface
sheet in MgB$_2$.

Fig.\ \ref{figamp} shows $F(k_\parallel)=\hbar A(k_\parallel)/2\pi e$ [where $A(k_\parallel)$ is
the Fermi surface cross-sectional area], calculated from LDA results, at several field angles. For
angles $\geq 65^\circ$, $F(k_\parallel)$ has two distinct extremal values. The contribution of this
finite Fermi surface region to the dHvA amplitude can be calculated by evaluating the integral
$I(B)=\int dk_\parallel \sin\left[\frac{2 \pi F(k_\parallel)}{B}\right]$ over a suitable range of
$k_\parallel$. The inset of Fig.\ \ref{figamp} shows the dHvA torque amplitude calculated in this
way, taking account of the angle dependence of $R_D$, $R_S$ and $\mathrm{d}F/\mathrm{d}\theta$. The
calculated amplitude has a series of sharp minima which resemble the single minimum seen in the
experimental data. The reason for the discrepancy in the number and angular location of the minima
is most likely due to a small inaccuracy of the LDA calculation: a warping approximately 30~T
smaller than that predicted by the LDA (at $\theta=90^\circ$) would place the first minimum at the
experimentally observed angle (other minima would be shifted to angles where strong damping would
prevent their observation). This level of error in the LDA calculation is entirely plausible
because the value predicted for $F_3$ is $\sim 200$~T too high before applying a rigid energy
shift. The absence of a true spin-zero for $\theta\gtrsim 50^\circ$ implies a Stoner enhancement $S
< 0.22$, in agreement with other estimates.

\section{Damping in the superconducting state}
\begin{figure}
\center
\includegraphics[width=6cm]{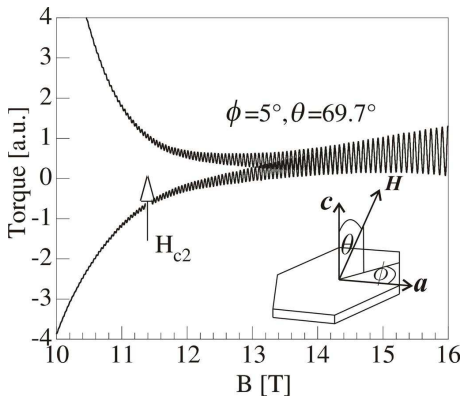}
\caption{Torque measured near $T_c$ with $\theta=69.7^\circ$ and $\phi\simeq 5^\circ$. At this
angle $H_{c2}\approx 11.5$\,T, as estimated from the rest of the torque loop. The substantial
irreversibility above this field is probably due to surface superconductivity. Note that this is
part of a minor hysteresis loop; in large samples the torque arising from superconductivity is too
large to measure over a complete loop.} \label{FIG:dhvascstate}
\end{figure}

In strongly type II, high $H_{c2}$, superconductors,  quantum oscillations can often be observed in
the superconducting state, i.e.,  at fields where superconducting vortices penetrate the sample
(see for example, Ref. \cite{JanssenHHMSW98}). In pure MgB$_2$, $\mu_0H_{c2}\sim 18$~T for fields
oriented close to the boron planes, and for field angles $\theta\gtrsim 65^\circ$ we are able to
observe dHvA oscillations from the $\pi$ band orbit $F_3$ below $H_{c2}$ \cite{FletcherCKK03}.
Fig.\,\ref{FIG:dhvascstate} shows an example of torque data obtained close to $H_{c2}$. Quantum
oscillations are superimposed on the background torque due to superconductivity, which arises from
the anisotropy of the coherence length and pinning \cite{KoganPRB88}. As is generally found in
other materials, the dHvA frequency remains identical to that in the normal state but the
oscillations suffer an additional attenuation beyond the factors accounted for in the conventional
Lifshitz-Kosevich treatment.

The survival of quantum oscillations deep into the superconducting state is surprising for several
reasons: (i) the vortex state is characterized by an intrinsic field inhomogeneity on the scale of
the London penetration depth, (ii) pinning of vortices leads to a vortex density gradient and thus
field inhomogeneity on the scale of the sample, and (iii) the opening of the superconducting gap
modifies the low energy quasiparticle spectrum, reducing the number of carriers available to
undergo cyclotron motion. Despite this, oscillations can sometimes be seen at fields as low as
0.2~$H_{c2}$ \cite{HeineckeW95}. A great deal of theoretical work \cite{YasuiK02} (see
Ref.~\cite{ManivRMP01} for a review) has been performed to establish whether the origin of the
oscillations is the same as that in the normal state, and to understand the mechanisms which
attenuate the oscillations. Many models attribute the attenuation to the reduction in the
(spatially averaged) density of states at the Fermi energy in the vortex state\cite{BrandtPT67},
which is related to the orbit-averaged superconducting gap. This is particularly interesting as the
size and $\bm{k}$-dependence of the additional damping might be used to determine the
$\bm{k}$-dependence of the superconducting gap. Rigorous experimental tests of the various
theoretical models for damping of quantum oscillations in the superconducting state have been hard
to perform because such observations are generally made in exotic materials where even the normal
state is not well understood: in some cases only a fraction of Fermi surface is detected (even in
the normal state) and the nature of the effective mass renormalization is unknown. Our knowledge of
the electronic structure and superconducting properties of MgB$_2$,  make this an ideal material
for performing such a study.

\begin{figure}
\center
\includegraphics[width=6cm]{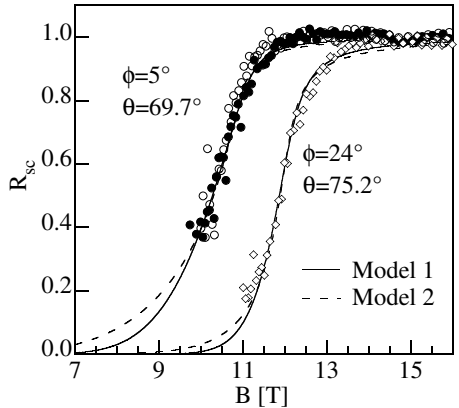}
\caption{Additional attenuation of the oscillatory torque signal in MgB$_2$ in the superconducting
(sc) state. For $\theta=69.7^\circ$, data taken on up sweeps ($\circ$) and down sweeps ($\bullet$)
are shown. Data are shown for in-plane rotations $\phi=5^\circ$ and $\phi=24^\circ$. Due to fine
details of the $\pi$-band Fermi surface sheet (see text), measurements at $\phi=24^\circ$ allow
$R_\mathrm{sc}$ to be determined over a wider field range.} \label{FIG:rs}
\end{figure}

A common way of isolating the additional attenuation is by dividing out other field dependent
factors from the original LK expression (see Ref.~\cite{FletcherCKK03} for details). This
additional damping factor, $R_{\rm sc}$, is shown in Fig.~\ref{FIG:rs} for MgB$_2$. The results
were found to be similar at all measured angles, apart from changing the field values at which the
attenuation begins. The attenuation has a rather rounded onset, but is quite severe, corresponding
to an order of magnitude, at 0.9~$H_{c2}$.

The value at which attenuation becomes significant is substantially lower than the point at which
the torque becomes irreversible. In fact, as shown in Fig.~\ref{FIG:rs}, irreversibility extends to
much larger fields than is expected from the rest of the superconducting torque loop. This is most
likely due to surface superconductivity \cite{Isshiki2005,RydhWHKKCKKJLKL03}, although some
calculations suggest that fluctuation effects may be important in this region \cite{Maniv2006}.

The field dependence of $R_{sc}$ was found to be independent of sweep direction, i.e., unaffected
by the irreversibility of the background torque, and was reproduced in other samples with different
degrees of vortex pinning. This suggests that the signal is unaffected by inhomogeneity of the the
vortex distribution due to pinning.  Following the arguments in Ref.\ \cite{JanssenHHMSW98}, it is
clear that the small scale field inhomogeneity due to a homogenous vortex lattice also does not
produce significant attenuation of the dHvA signal in MgB$_2$.

In another study of this effect in MgB$_2$ \cite{Isshiki2005}, an ac field was superimposed on the
main field and modulations in torque were detected. If vortices are pinned at the sample surfaces,
the ac field modulation does not penetrate the bulk of the sample and an additional damping of the
observed signal results. The dc field torque method used in the work described here
\cite{FletcherCKK03} is not affected by this problem.

The damping of the dHvA signal is closely linked with the onset of superconductivity in the bulk of
the sample. In the present case,  additional attenuation seen in the dHvA signal is attributed to
the finite gap present on the $\pi$ band, which is present even in fields close to $H_{c2}$. Very
few probes of the superconducting gap exist in such high fields, and in fact it was previously
suggested that the superconductivity in the $\pi$ bands was quenched at much lower fields
\cite{DagheroGUSJKK03}.

In Fig.~\ref{FIG:rs} two models are applied to the measured damping. The zero field superconducting
gap $\Delta_0$ and $H_{c2}$ are free parameters. An additional parameter is included to account for
the broadening of the transition \cite{ClaytonIHMSS02}. The first model \cite{Maki91} is based on
the calculation of the average density of states in the vortex lattice \cite{BrandtPT67}. This
requires $\Delta_0\sim$200-300~K in order to describe the data. An alternative model
\cite{Miyake93}, simply based on the zero field (and hence fully gapped) density of states,
requires a value of $\Delta_0\sim$ 30-60~K, closer to that expected in MgB$_2$. Compared to results
in other materials\cite{JanssenHHMSW98} this level of agreement is not unusual. While this may
indicate that this effect has yet to be fully understood, we also note that there are some
limitations in this analysis. Firstly, data are only available at high fields, requiring a
substantial extrapolation of the field dependence of the gap back to the zero field value. This
field dependence is non-trivial in the multi-band case \cite{GraserDS04}. In addition, the meaning
of the gap parameter in each model can be different, i.e., it represents an effective gap, which
may be quantitatively different from the zero field value. In the light of this, it would be
interesting to observe dHvA signals below $H_{c2}$ on both $\sigma$ and $\pi$ bands in MgB$_2$ to
confirm that the relevant parameter is the sheet-dependent (orbit-averaged) gap. This would require
the dHvA signal to survive down to fields of order $\mu_0H_{c2}^{\|c}\simeq 3.5$T.  We estimate
that this would require samples with $\sigma$ sheet mean-free-paths approximately three times
larger than we measured to date \cite{FletcherCKK03}.

\section{Conclusions}
In conclusion, we have seen that in general, our measurements are in excellent agreement with the
predictions of the LDA band structure, including the strength of the electron-phonon coupling on
each Fermi surface sheet.  The minor discrepancies which remain can be explained with small rigid
band shifts. Our work on Al substitution confirms that adding Al dopes electrons into the
structure, and reduces the size of the $\sigma$ sheets. The size of the reduction and the size of
the $T_c$ decrease is \textit{quantitatively} explained by the VCA calculations.

Although our understanding of the dHvA data obtained until now is rather good, the data for Al
doped samples show that there is scope for further studies of crystals lightly doped with other
elements. These would give important insight as to how the mean-free-paths on various orbits and
the strengths of the electron-phonon coupling can be altered in a controlled manner with alloying
and give deeper understanding as to how $T_c$ and $H_{c2}$ can be optimized for applications.

The dHvA effect in the superconducting state, however, is still not fully understood.  The
existence of two distinct gaps in MgB$_2$ gives us an important new parameter with which to study
this effect - especially if in the future higher purity samples become available so that dHvA
oscillations can be measured at fields less than $\mu_0H_{c2}^{\|c}\simeq 3.5$T and the
superconducting gap on the $\sigma$ sheet can also be probed.

\section{Acknowledgements}
High quality single crystal samples were provided by J.~Karpinski, S.M.~Kazakov and N.D.~Zhigadlo
of ETH Z\"{u}rich (pure, Al and C doped MgB$_2$), and S.~Lee, A.~Yamamoto, and S.~Tajima of
University of Tokyo (pure MgB$_2$).  X-ray characterization of the Al-doped samples was performed
by J.P.H.~Charmant.  Important contributions to the dHvA experiments were also made by P.J.~Meeson,
N.E.~Hussey, and L.~Balicas. Theoretical work linking the LDA calculations to the dHvA experiments
was done by J.~Kortus , I.I.~Mazin , and H.~Harima. Work in the UK was supported by the EPSRC, in
Japan by NEDO and in Florida by the NSF.


\begin{thebibliography}{10}
\bibitem{stonernote}
These factors are slightly higher than the ones reported in Ref.\
  \cite{CarringtonFH05} ($S(\mathrm{F_1})=0.07$ and $S(\mathrm{F_2})=0.12$).
  The difference arises because in Ref.\ \cite{CarringtonFH05} we used the raw
  bandstructure values of $m_B(\theta)$ whereas here we have used the values
  calculated from the bands which have been rigidly shifted to agree with the
  $F(\theta)$ data. This shows the sensitivity of determination of $S$ to
  uncertainties in the bandstruture - particularly for high angles.
\end{thebibliography}

\end{document}